\documentclass[12pt]{article}
\usepackage{amssymb,amsmath,cite}
\usepackage{epsfig}
\long\def\@makefntext#1{\parindent 0cm\noindent \hbox to
1em{\hss$^{\@thefnmark}$}#1}

\begin{document}
\begin{titlepage}
\vspace{.5in}
\begin{flushright}
\end{flushright}
\vspace{.5in}
\begin{center}
{\Large\bf   Non-radial oscillations   of anisotropic neutron stars in the Cowling approximation}\\
\vspace{.4in} {Daniela D. Doneva$^{1,3}$\footnote{\it email:
daniela.doneva@uni-tuebingen.de},
               Stoytcho~ S.~Yazadjiev$^{1,2}$\footnote{\it email: yazad@phys.uni-sofia.bg}\\
       { \footnotesize  ${}^{1}$ \it Theoretical Astrophysics, IAAT, Eberhard-Karls University of T\"ubingen, T\"ubingen 72076, Germany }}\\
       {\footnotesize\it ${}^{2}$ \it Department of Theoretical Physics, Faculty of Physics,}
       {\footnotesize \it Sofia University, Sofia, 1164, Bulgaria } \\
       {\footnotesize \it ${}^{3}$ Institute for Nuclear Research and Nuclear Energy, Bulgarian Academy of Sciences, Sofia, Bulgaria}
\end{center}

\vspace{.5in}

\begin{center}
{\large\bf Abstract}
\end{center}

 One of the most common assumptions in the study of neutron star models and their oscillations is that
the pressure is isotopic, however there are arguments that this may not
be correct. Thus in the present paper we make a first step towards
studying the nonradial oscillations of neutron stars with an
anisotropic pressure. We adopt the so-called Cowling approximation
where the spacetime metric is kept fixed and the oscillation
spectrum for the first few fluid modes is obtained. The effect of
the anisotropy on the frequencies is apparent, although with the
present results it might be
hard to distinguish it from the changes in the frequencies caused by different equations of state. \\ \, \\

PACS:  04.40.Dg;  04.30.Db
\end{titlepage}
\addtocounter{footnote}{-1}

\section{Introduction}
The discovery of gravitational waves is one of the most important
goals of the astrophysics nowadays. A lot of effort is being devoted to
this problem worldwide. Ground based experiments
\cite{LIGO}--\cite{VIRGO2} as well as space missions \cite{LISA} are
planned and some of them are expected to be able to give results in
the near future. Parallel to these efforts the first steps towards a
third generation gravitational wave telescope (the so-called
Einstein telescope) which is supposed to  have much higher
sensitivity are being undertaken \cite{EinsteinTel}. The reason why the
gravitational waves are so difficult to detect is that they are
extremely weak which requires detectors with very high sensitivity
and also accurate waveforms of the signal emitted from the
astrophysical objects.

One of the promising sources of gravitational waves is the
oscillations of neutron stars \cite{KokkRev}. A lot of effort has been
spent in studying the gravitational wave emission of these objects
but there is still many important unanswered questions. The
gravitational wave emission by neutron stars is ultimately connected
to their interior structure. In order to predict  accurately enough
the characteristics of the gravitation waves we need adequate
relativistic models of the neutron star interior. However, at
present  little is known about the properties and the behaviour of
matter at very high densities and pressures. So, in modeling the
neutron star interior we are forced to make certain assumptions
about the properties of the neutron star matter. Some of these
assumptions seem natural from a physical point of view,
however, there are always uncertainties and suspicions that the
assumptions may be not fully correct. As the
history of science shows, there are surprises sometimes -- Nature does not always share our
notions of what is  ``natural''. That is why the alternatives  should also be investigated.

One of the widely accepted assumptions in studying the equilibrium configurations of neutron stars and their
oscillations is that the pressure of the neutron star matter is isotropic.  There are however arguments that the
pressure could be anisotropic\footnote{ Generally speaking the anisotropic fluid has pressures which can differ among
the spacial directions.} \cite{Herrera}. Some theoretical investigations \cite{Ruderman,Canuto74} show that the nuclear
matter may be anisotropic at very high densities where the nuclear interactions must be treated relativistically.
Anisotropy in the fluid pressure can be caused by many other factors. Anisotropy can be yielded by the existence of a
solid core or by the presence of superfluid \cite{KW}--\cite{Heiselberg00}, by pion condensation \cite{Sawyer}, by
different kind of phase transitions \cite{Sokolov}, by the presence of strong magnetic field \cite{Y} or by other
factors \cite{Herrera}. From a formal point of view the mixture of two fluids is mathematically equivalent to an
anisotropic fluid \cite{Herrera},\cite{L}.

During the last decades, starting with the pioneering work of \cite{BL},
there have been many papers studying anisotropic spherically
symmetric static  configurations within general relativity
\cite{HH}--\cite{BI1}. These studies show that the anisotropy may
have non-negligible effects on the neutron star structure and
properties. For example the anisotropy may influence notably the
maximal equilibrium mass, maximum redshift and maximum compactness
of the stars \cite{HH}, \cite{BI}. It is worth noting also that even
for stable configurations the anisotropy can support outwardly
increasing energy density in the star core \cite{Horvat}.

The fact that the anisotropy can seriously affect the interior
structure, and the properties of the stellar configurations make us
think that the anisotropy may also  have a serious influence on the
gravitational wave emission and more precisely on the gravitational
wave spectrum of the stellar configurations. Therefore, in the
context of the current efforts to detect the gravitational waves, it
is important to study the gravitational wave spectrum of the
anisotropic neutron stars. Such a study is twofold. On the one hand, it
can reveal the basic characteristics of the gravitational wave
spectrum of the anisotropic stars and the differences with the
spectrum of the isotropic stars. On the other hand, such a study
provides us with a tool to study the reverse problem -- to put
constraints on the amount of neutron star anisotropy using the
observed gravitational wave spectrum in the future.

In the present paper we undertake the first step towards the study
of the gravitational wave spectrum of anisotropic neutron stars.
More precisely, we investigate the spectrum of the nonradial
oscillations of  anisotropic neutron stars in the Cowling approximation.
The paper is organized as follows. In section 2 we numerically
construct equilibrium configurations describing anisotropic stars.
Section 3 is devoted to the derivation of the perturbation equations
of the anisotropic neutron stars in the Cowling approximation and the
formulation of the boundary value problem for the oscillation
spectrum. In section 4 we present the numerical results for the
oscillation frequencies. The paper ends with conclusions.

\section{Equilibrium anisotropic configurations of neutron stars}

In the spherically symmetric case\footnote{The assumption of
spherical symmetry is applicable to the static case of matter
sources with an energy-momentum tensor satisfying $|
T^{\theta}_{\theta}-T^{\phi}_{\phi}|<< T^{\theta}_{\theta}$. So the
assumption of spherical symmetry is applicable, for example,  to the
cases when the anisotropy is yielded by the existence of a solid
core, by the presence of superfluid, or by the presence of pion
condensation. Also, the spherical symmetry can be used when the
anisotropy is yielded by weak enough magnetic field. However, if the
magnetic field is very strong, as in the case of magnetars, the
spherical symmetry assumption may not be a good
approximation\cite{Y}. } which we will consider in the present
paper, the fluid anisotropy means that the radial pressure $p$
differs from the transverse pressure $q$. The mathematical
description of an anisotropic fluid in spherical symmetry is given
by the following energy-momentum tensor

\begin{eqnarray}
T_{\mu\nu}=\rho u_{\mu} u_{\nu} + p k_{\mu} k_{\nu} + q
\left(g_{\mu\nu} + u_{\mu} u_{\nu} - k_{\mu} k_{\nu} \right),
\end{eqnarray}
where $g_{\mu\nu}$ is the spacetime metric, $u^{\mu}$ is the fluid 4-velocity, $\rho$ is the fluid energy density and
$k^{\mu}$ is the unit radial vector ($k_{\mu}k^{\mu}=1$) with $u^{\mu}k_{\mu}=0$. Note that $g_{\mu\nu} + u_{\mu}
u_{\nu} - k_{\mu} k_{\nu}$ is the projection tensor onto the 2-surfaces orthogonal to both $u^{\mu}$ and $k^{\mu}$. At
the center of symmetry the anisotropic pressure must vanish since $k^{\mu}$ is  not defined there.

For  spherically symmetric spacetimes the  metric can be written in
the well-known form

\begin{eqnarray}
ds^2 = - e^{2\Phi}dt^2 + e^{2\Lambda}dr^2 + r^2d\theta^2 + r^2
\sin^2\theta d\phi^2.
\end{eqnarray}
The Einstein field equations
\begin{eqnarray}
R_{\mu\nu} - \frac{1}{2}g_{\mu\nu} R= 8\pi T_{\mu\nu}
\end{eqnarray}
then reduce to
\begin{eqnarray}\label{LEQ}
&&\frac{2\Lambda^\prime}{r} e^{-2\Lambda} + \frac{1}{r^2}\left(1-
e^{-2\Lambda}\right)= 8\pi \rho ,\\
&&\frac{2\Phi^\prime}{r} e^{-2\Lambda} - \frac{1}{r^2}\left(1-
e^{-2\Lambda}\right)=8\pi p \label{PHIEQ},
\end{eqnarray}
while the contracted Bianchi identity

\begin{eqnarray}\label{Bianchi}
\nabla^{\mu}T_{\mu\nu}=0
\end{eqnarray}
gives

\begin{eqnarray}\label{FEEQ}
&&p^\prime = - (\rho + p)\Phi^{\prime} - \frac{2\sigma}{r},
\end{eqnarray}
where $\sigma=p-q$.  Introducing the local mass
$m(r)=\frac{r}{2}\left(1- e^{-2\Lambda}\right)$ and expressing
$\Phi^\prime$ from (\ref{PHIEQ}) we can write the dimensionally
reduced equations in the Tolman-Oppenheimer-Volkoff form

\begin{eqnarray}
&&m^{\prime}= 4\pi \rho r^2, \\
&& p^{\,\prime}= - (\rho+ p)\, \frac{4\pi pr^3 + m}{r\left(r -
2m\right)} - \frac{2\sigma}{r}.
\end{eqnarray}

In order to close our system we should specify the equations of state for $p$ and for $\sigma$. For the radial pressure
we will consider a barotropic equation  of state and more precisely

\begin{eqnarray}
\rho= p_0 \left(\frac{p}{Kp_0}\right)^{1/\Gamma} + \frac{p}{\Gamma -1},
\end{eqnarray}
where $p_0=1.67 \times 10^{14}~\rm{g/cm^3}$ in units where $c=1$ and
we have chosen $\Gamma=2.34$ and $K=0.0195$ obtained when fitting
the tabulated data for EOS II \cite{Alonso85}. The results are
qualitatively the same for other values of  $K$ and $\Gamma$.

As explained in \cite{Horvat} we cannot  just take
$\sigma=\sigma(\rho)$ because this equation of state is too
restrictive. Instead we should consider quasilocal equation of
state $\sigma=\sigma(p,\mu)$ where $\mu$ denotes a quasilocal
variable. In principle the equation of state
$\sigma=\sigma(\rho,\mu)$ should be determined by the microscopic
theory. Unfortunately, at present we do not have a good enough
microscopic theory to allow us to find the explicit form of the
dependence $\sigma=\sigma(\rho,\mu)$. A drawback of the available
microscopic models is the fact that the models are developed in flat
spacetime and then the results are transferred to curved spacetime
which is not completely satisfactory \cite{Glendenning97}. Within
the framework of these types of microscopic models  it is impossible to
find the influence of the curved geometry on the equation of state.
That is why our approach in the present paper is phenomenological.
Following \cite{Horvat} for the quasilocal variable we take the
local compactness $\mu=\frac{2m(r)}{r}=1-e^{-2\Lambda}$ and we
consider the following equation of state

\begin{eqnarray}
\sigma=\lambda p \mu,\label{eq:EOS_sigma}
\end{eqnarray}
where $\lambda$ is a parameter. Since the local compactness is zero
at the center, this guarantees that $\sigma(0)=0$. In order to
roughly estimate the range of the parameter $\lambda$ we use the
results of \cite{Sawyer1} where the anisotropy is caused by a pion
condensation. In \cite{Sawyer1} it is found that $0\le \sigma/p \le
1$ and therefore we could expect that the maximum  value of
$\lambda$ is of order of 1. In the present paper we adopt the range
$\lambda \in [-2,2]$.

In order to obtain the background solutions which will be perturbed we solve the reduced field equations
(\ref{LEQ}),(\ref{PHIEQ}) and (\ref{FEEQ}) with the appropriate boundary conditions
\begin{eqnarray}
\Lambda(0)=0,\quad \rho(0)=\rho_0, \quad \Phi(\infty)=0.
\end{eqnarray}

\begin{figure}
\begin{center}
\includegraphics[width=0.6\textwidth]{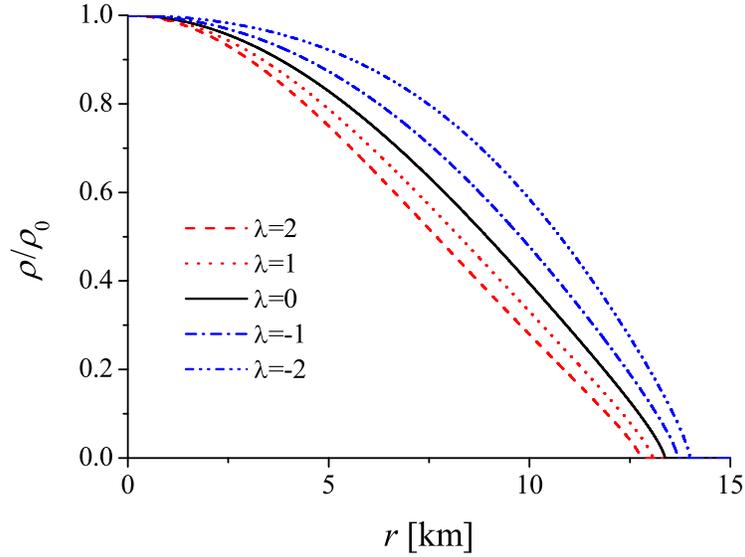}
\end{center}
\caption{The normalized energy density $\rho$ as a function of the radial coordinate~$r$. The results for several
neutron-star solutions with the same central energy density $\rho_0~=~7.455~\times~10^{14}~\rm{g/cm^3}$ and different
values of the parameter $\lambda$ are shown (this central energy density gives neutron star with mass $M=1.4
M_{\bigodot}$ in the case with zero anisotropic pressure, i.e. when $\lambda=0$).
} \label{fig:rho(x)}%
\end{figure}

\begin{figure}
\includegraphics[width=0.48\textwidth]{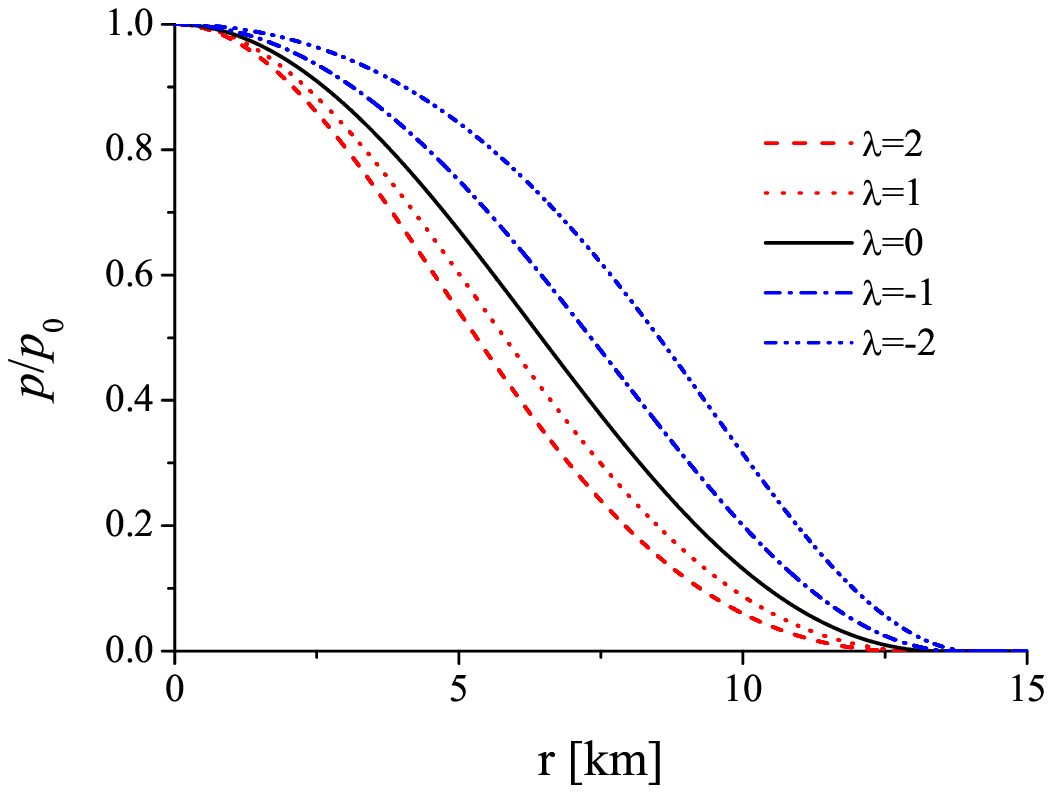}
\includegraphics[width=0.48\textwidth]{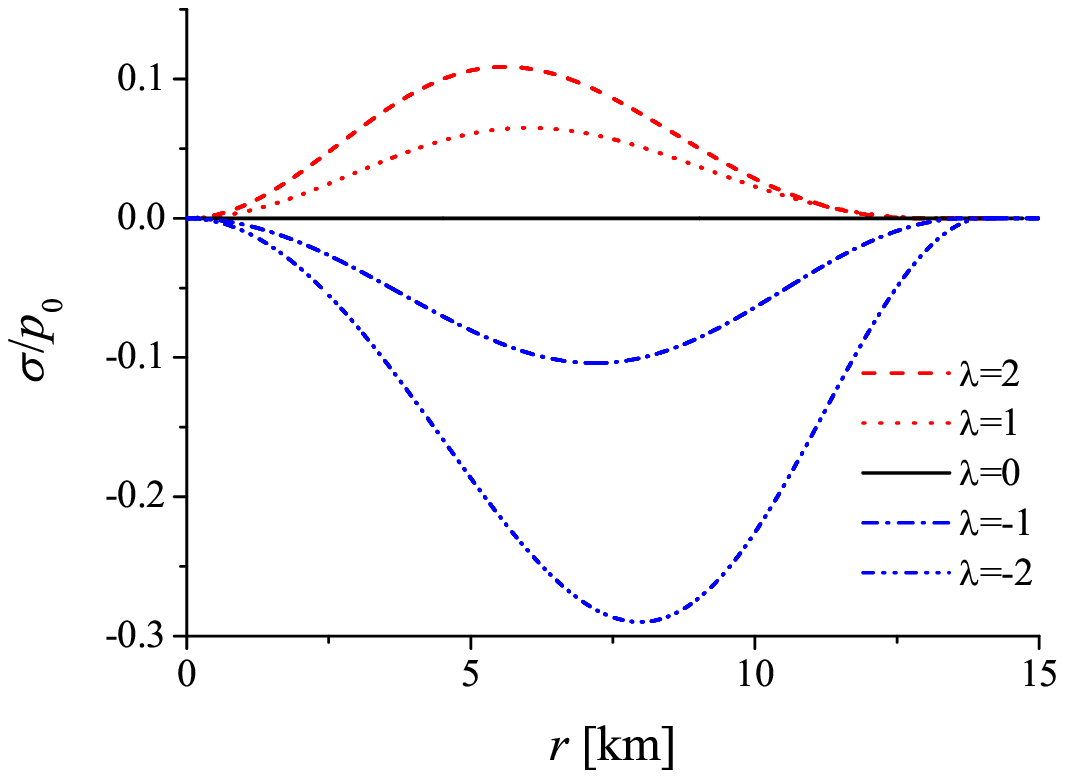}
\caption{The radial $p$ and the anisotropic pressure $\sigma$ as
functions of the radial coordinate $r$, normalized to the value of
the radial pressure at the center of the star $p_0$. The results are
for the same solutions as in Fig. \ref{fig:rho(x)}.
} \label{fig:p_sigma(x)}%
\end{figure}

The normalized density $\rho$, the radial $p$, and the anisotropic
$\sigma$ pressure as  functions of the radial coordinate $r$ are
shown in Figs. \ref{fig:rho(x)} and \ref{fig:p_sigma(x)} for several
neutron star solutions. The central energy density $\rho_0$ is the
same for all of the solutions presented in the figures and the
results for several values of the parameter $\lambda$, which
controls the anisotropic pressure, are shown. It is interesting to
note that for large values of $\lambda$ and for large masses,
neutron star solutions exist for which the density $\rho$ is not a
monotonic function of the radial coordinate but has a maximum and
these solutions are dynamically stable \cite{Horvat}.

\begin{figure}
\includegraphics[width=0.48\textwidth]{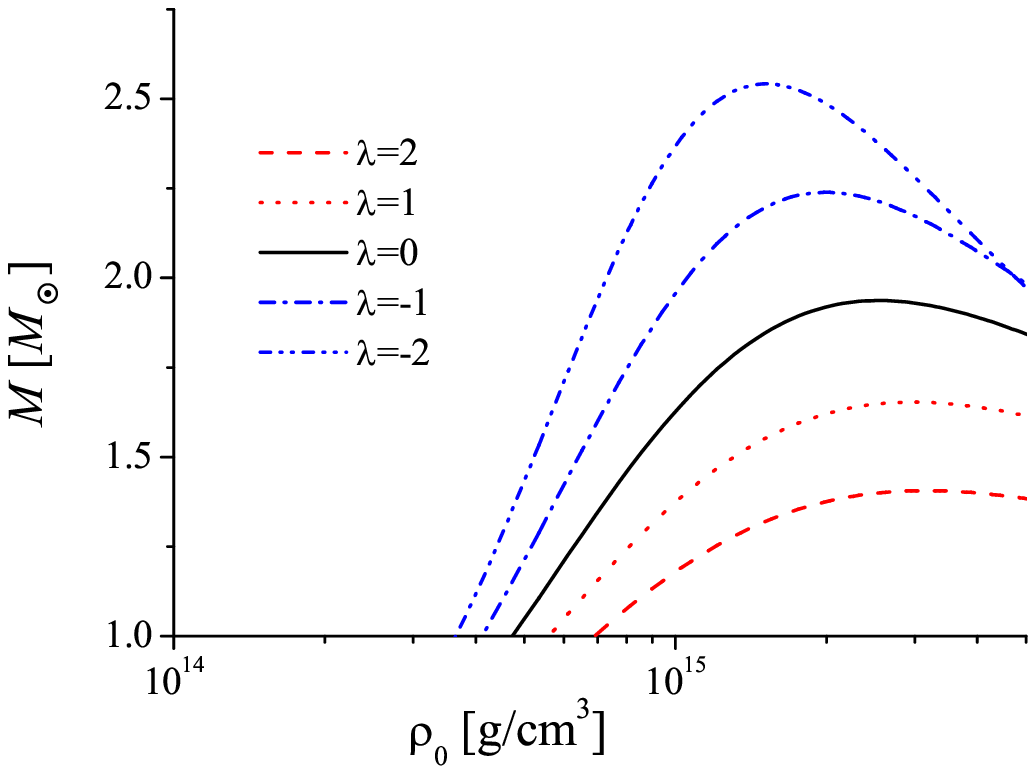}
\includegraphics[width=0.48\textwidth]{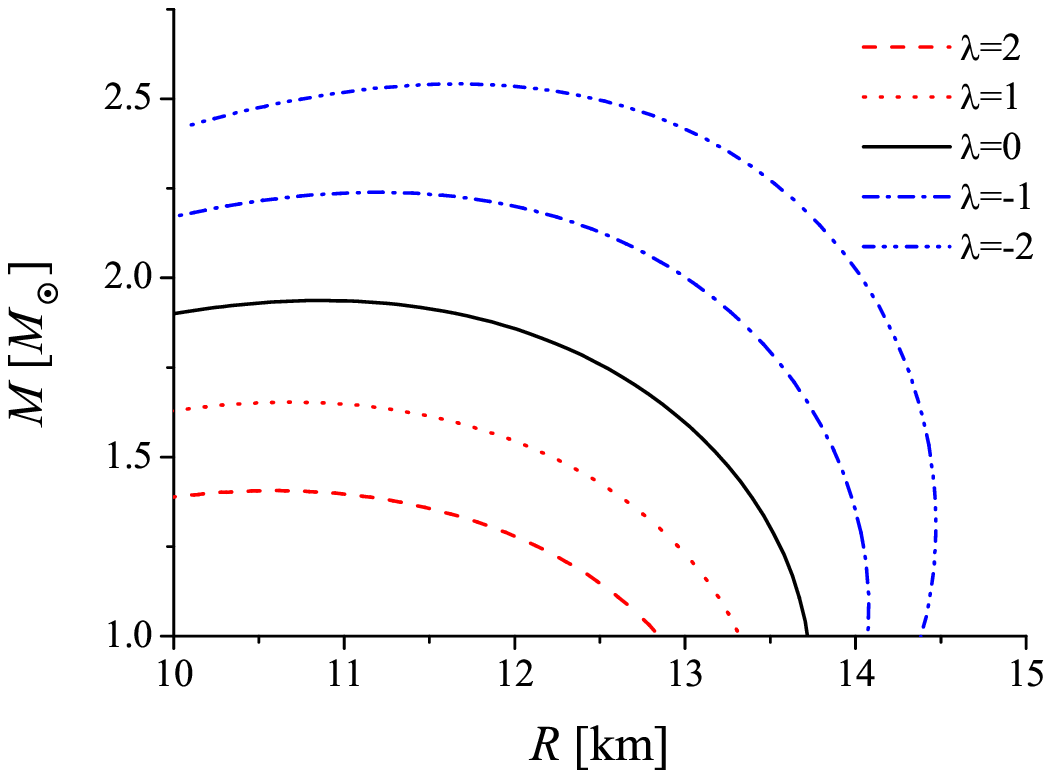}
\caption{The mass of the neutron stars as a function of the central density (left panel) and of the radius (right
panel) for several values of the parameter $\lambda$.}
\label{fig:M_(rho)(R)}%
\end{figure}

In Fig. \ref{fig:M_(rho)(R)} the mass $M$ of the anisotropic neutron
stars is shown as a function of the central density $\rho_0$ and of
the radius $R$ for several values of the parameter $\lambda$. As we
can see the properties of the star vary significantly when the
anisotropic pressure is varied, i.e. when we vary the parameter
$\lambda$. The dynamical stability analysis shows that the solutions
are stable up to the maximum mass of the sequences \cite{Horvat}.

\section{Perturbation equations in the Cowling approximation}

In this section we derive the equations describing the nonradial perturbations of the anisotropic stars in the
so-called Cowling approximation \cite{McDermott},\cite{Lindblom}. In the Cowling approximation the spacetime metric is kept
fixed.  Despite this simplification the Cowling formalism turns out to be accurate enough and reproduces the
oscillation spectrum with good accuracy. In fact, the comparison of the oscillation frequencies obtained by a fully
general relativistic numerical approach and by the Cowling approximation shows that the discrepancy is less than $20
\%$ for the typical stellar models \cite{Yoshida}.

The equations describing the perturbations in the Cowling formalism are obtained by varying the equations for the
conservation of the energy-momentum tensor (\ref{Bianchi}). Taking into account that the metric is kept fixed, we find
$\nabla_{\nu}\delta T^{\nu}_{\mu}=0$ where

\begin{eqnarray}
\delta T^\nu_{\mu}= \left(\delta \rho + \delta q\right) u_{\mu}
u^{\nu} + \left(\rho + q\right)\left(u_{\mu}\delta u^{\nu} + \delta
u_{\mu} u^{\nu} \right) + \delta q\delta^{\,\nu}_{\mu} +
\delta\sigma k^{\nu}k_{\mu} + \sigma \delta k^{\nu} k_{\mu} + \sigma
k^{\nu} \delta k_{\mu}.
\end{eqnarray}

Projecting  equation $\nabla_{\nu}\delta T^{\nu}_{\mu}=0$  along the
background 4-velocity $u^{\mu}$  we have

\begin{eqnarray}\label{PPU}
u^{\nu}\nabla_{\nu}\delta\rho + \nabla_{\nu}\left\{\left[(\rho +
q)\delta^{\nu}_{\mu}+ \sigma k^{\nu}k_{\mu}\right]\delta
u^{\mu}\right\} + \left(\rho + q\right)a_{\nu}\delta u^{\nu} +
\nabla_{\nu}u_{\mu}\delta \,\left(\sigma k^{\nu}k^{\mu}\right)=0.
\end{eqnarray}

Projecting orthogonally to the background 4-velocity by using the
operator ${\cal P}_{\mu}^{\nu}=\delta^{\nu}_{\mu} + u^{\nu}u_{\mu}$,
we obtain

\begin{eqnarray}\label{POE}
\left(\delta\rho + \delta q\right)a_{\mu} + \left(\rho +
q\right)u^{\nu} \left(\nabla_{\nu}\delta u_{\mu} -
\nabla_{\mu}\delta u_{\nu} \right) + \nabla_{\mu}\delta q + u_{\mu}
u^{\nu}\nabla_{\nu}\delta q + {\cal
P}_{\mu}^{\nu}\nabla_{\alpha}\delta\left(\sigma k^{\alpha }k_{\nu}
\right) =0,
\end{eqnarray}
where $a_{\mu}=u^{\nu}\nabla_{\nu}u_{\mu}$ is the background
4-acceleration.

At this stage we can express the perturbations of the 4-velocity via
the Lagrangian displacement vector $\xi^{i}$, namely,

\begin{eqnarray}
\frac{\partial \xi^{i}}{\partial t}= \frac{\delta u^{i}}{u^t},
\end{eqnarray}
where $i=1,2,3=r,\theta,\phi$.

Now let us consider  eq.(\ref{POE}) for $\mu=\theta$ and $\mu=\phi$.
Since $a_{\theta}=a_{\phi}=0$ and $u^\mu=(u^{t},0,0,0) $ we find

\begin{eqnarray}
(\rho + q)(u^{t})^2\partial^2_{t}\xi_{\theta}+
\partial_{\theta}\delta q=0, \\ \nonumber \\
(\rho + q)(u^{t})^2\partial^2_{t}\xi_{\phi}+
\partial_{\phi}\delta q=0.
\end{eqnarray}
Taking into account that $\rho, q$ and $u^{t}$ depend on $r$ only,
the integrability condition for the above equations gives

\begin{eqnarray}
\partial_{\theta}\xi_{\phi}= \partial_{\phi}\xi_{\theta}.
\end{eqnarray}
From this condition and the fact that the background is spherically
symmetric we find that $\xi_{\theta}$ and $\xi_{\phi}$ are of the
form

\begin{eqnarray}
\xi_{\theta}= -\sum_{lm}V_{lm}(r,t)\partial_{\theta}Y_{lm}(\theta,\phi),\\ \nonumber\\
\xi_{\phi}= -\sum_{lm}V_{lm}(r,t)\partial_{\phi}Y_{lm}(\theta,\phi),
\end{eqnarray}
where $Y_{lm}(\theta,\phi)$ are the spherical harmonics. From now on, in order to simplify the notations, we will just
write $\xi=-V Y_{lm}$ when we have expansion in spherical harmonics.

We proceed further with finding the expressions for the density and
pressure  perturbations. From eq.(\ref{PPU}) after some algebra we
find

\begin{eqnarray}
\delta \rho = -
\frac{1}{\sqrt{-g}}\partial_{i}\left\{\sqrt{-g}\left[(\rho
+q)\xi^{i} + \sigma (k_j\xi^{j})k^{i}\right]\right\}- \nonumber\\
\left[(\rho+q)\xi^{i}+
\sigma(k_{j}\xi^{j})k^{i}\right]\partial_{i}\ln(u^{t}) - (\rho +
p)a_{i}\xi^{i}.
\end{eqnarray}
It is convenient to express $\xi^{r}$ in the form
\begin{eqnarray}
\xi^{r}= e^{-\Lambda} \frac{W}{r^2} Y_{lm}
\end{eqnarray}
 and substituting in the above equations, after some algebra
we find\footnote{The derivative with respect to the radial
coordinate $r$ will be denoted by prime or by the standard symbol
interchangeably.}

\begin{eqnarray}
\delta \rho = - (\rho + p) \left[e^{-\Lambda} \frac{W^\prime}{r^2} +
\frac{l(l+1)}{r^2} V \right] Y_{lm} -
\frac{d\rho}{dr}e^{-\Lambda}\frac{W}{r^2} Y_{lm} +
\frac{2\sigma}{r^3}e^{-\Lambda}W Y_{lm} + \sigma
\frac{l(l+1)}{r^2}VY_{lm},
\end{eqnarray}
where in the last step we have taken into account that
$a_{r}=\Phi^{\prime}$ and eq. (\ref{FEEQ}).

In order to find the perturbation of the radial pressure we first
use the relation between the Eulerian and Lagrangian variations,
namely

\begin{eqnarray}
\delta p = \Delta p - \xi^{r} \partial_{r}p
\end{eqnarray}
with $\Delta p$ being the Lagrangian variation. From the equation of
state we have

\begin{eqnarray}
\Delta p= \frac{dp}{d\rho}\Delta \rho= \frac{dp}{d\rho} \left(\delta
\rho + \xi^{r}\partial_{r}\rho\right).
\end{eqnarray}
In this way we obtain the following formula for the perturbation of
the radial pressure

\begin{eqnarray}
\delta p=-\frac{dp}{d\rho} \left\{(\rho +
p)\left[e^{-\Lambda}\frac{W^\prime}{r^2} +
\frac{l(l+1)}{r^2}V\right] - 2\frac{\sigma}{r^3}e^{-\Lambda}W -
\sigma \frac{l(l+1)}{r^2}V\right\}Y_{lm} -
\frac{dp}{dr}e^{-\Lambda}\frac{W}{r^2}Y_{lm}.
\end{eqnarray}

For the perturbation of the anisotropic pressure
$\sigma=\sigma(p,\mu)$ we have

\begin{eqnarray}
\delta \sigma= \frac{\partial\sigma}{\partial p}\delta p,
\end{eqnarray}
where we have taken into account that $\delta \mu=0$.

The dynamical equations for $W$ and $V$  follow from eq. (\ref{POE}), namely

\begin{eqnarray}
&&(\rho + p)\frac{e^{\Lambda - 2\Phi}}{r^2} \partial^2_{t} W +
\partial_{r} {\hat {\delta p}}  + (\hat {\delta\rho} + \hat {\delta p})a_{r} +
\frac{2}{r}\hat {\delta \sigma}=0, \\
&&(\rho + p -\sigma) e^{-2\Phi}\partial^2_{t} V  - \hat {\delta p} +
\hat{\delta\sigma}=0,
\end{eqnarray}
where $\hat {\delta p}$  are the coefficients in the expansion in
the spherical harmonics $Y_{lm}$, i.e $\delta p = \hat {\delta p}
Y_{lm}$.

From now on we will assume for the perturbation functions a harmonic
dependence on time, i.e. $W(r,t)=W(r)e^{i\omega t}$ and
$V(r,t)=V(r)e^{i\omega t}$. Then the above equations become

\begin{eqnarray}\label{EQW}
&& - (\rho + p)\frac{e^{\Lambda - 2\Phi}}{r^2} \omega^2 W +
\partial_{r} {\hat \delta p}  + (\hat{\delta\rho} + {\hat \delta p})a_{r} +
\frac{2}{r}\hat{\delta \sigma}=0, \\
&&-(\rho + p -\sigma) e^{-2\Phi}\omega^2 V  - \hat {\delta p} +
{\hat \delta\sigma}=0 \label{EQV}.
\end{eqnarray}

The system (\ref{EQW})--(\ref{EQV}) can be considerably simplified
by combining the equations in an appropriate manner. Differentiating
equation (\ref{EQV}) and adding it to equation (\ref{EQW}), and also
using eq.(\ref{FEEQ}), we find

\begin{eqnarray}
&&V^\prime = 2V \Phi^{\prime} - \left(1 -\frac{\partial
\sigma}{\partial p} \right) \frac{\rho + p}{\rho + p-\sigma}
\frac{e^{\Lambda}}{r^2} W \\
 && + \left[\frac{\sigma^\prime}{\rho +
p -\sigma} + \frac{\frac{d\rho}{dp}+ 1}{\rho + p-\sigma}\sigma \left(\Phi^\prime + \frac{2}{r}\right)  -
\frac{2}{r}\frac{\partial\sigma}{\partial p} - \left(1 - \frac{\partial\sigma}{\partial
p}\right)^{-1}\left(\frac{\partial^{2}\sigma}{\partial p^2} p^\prime + \frac{\partial^2\sigma }{\partial p\partial
\mu}\mu^\prime\right)\right] V. \nonumber
\end{eqnarray}
This equation together with  equation (\ref{EQW}) solved for
$W^\prime$, form a system which is equivalent to
(\ref{EQW})--(\ref{EQV}) but much more tractable:

\begin{eqnarray}
W^\prime&=& \frac{d\rho}{dp}\left[\omega^2 \frac{\rho + p-\sigma}{\rho + p} \left(1 - \frac{\partial\sigma}{\partial
p}\right)^{-1}e^{\Lambda - 2\Phi} r^2 V  + \Phi^{\prime} W\right] - l(l+1)e^{\Lambda} V  \label{eq:CowlingW}  \\ \nonumber \\
& +& \frac{\sigma}{\rho + p}\left[\frac{2}{r}\left(1+ \frac{d\rho}{dp}\right)W + l(l+1)e^{\Lambda}V \right],  \nonumber
\\ \nonumber \\
V^\prime &=& 2V \Phi^{\prime} - \left(1 -\frac{\partial \sigma}{\partial p} \right) \frac{\rho + p}{\rho + p-\sigma}
\frac{e^{\Lambda}}{r^2} W  \label{eq:CowlingV}\\ \nonumber \\
  &+& \Big[\frac{\sigma^{\,\prime}}{\rho +
p -\sigma} + \left(\frac{d\rho}{dp}+ 1\right)\frac{\sigma}{\rho + p-\sigma} \left(\Phi^\prime + \frac{2}{r}\right)
\nonumber \\ \nonumber \\
&-& \frac{2}{r}\frac{\partial\sigma}{\partial p} - \left(1 - \frac{\partial\sigma}{\partial
p}\right)^{-1}\left(\frac{\partial^{2}\sigma}{\partial p^2} p^\prime + \frac{\partial^2\sigma }{\partial p\partial
\mu}\mu^\prime\right)\Big] V .\nonumber
\end{eqnarray}

The boundary condition at  the star surface is that the Lagrangian
perturbation of the radial pressure vanishes

\begin{eqnarray}\label{FBC}
\omega^2 \frac{\rho + p - \sigma}{\rho + p}\left( 1-\frac{\partial\sigma}{\partial p}\right)^{-1}e^{-2\Phi}V +
\left(\Phi^\prime  + \frac{2}{r} \frac{\sigma}{\rho + p}\right)e^{-\Lambda}\frac{W}{r^2}=0.
\end{eqnarray}

The boundary conditions at the star center can be obtained by
examining the behaviour in the vicinity of $r=0$. For this purpose
it is convenient to introduce the new functions $\tilde W$ and
$\tilde V$ defined by

\begin{eqnarray}
W= \tilde W r^{l+1}, \;\;\; V=\tilde V r^l .
\end{eqnarray}

Then one can show that at $r=0$ the following boundary condition is
satisfied

\begin{eqnarray}\label{FCAC}
\tilde W= -l \tilde V.
\end{eqnarray}

\section{Oscillation spectrum of the anisotropic neutron stars}
The oscillation spectrum of the anisotropic neutron stars in Cowling
approximation can be obtained by solving the differential equations
(\ref{eq:CowlingW})--(\ref{eq:CowlingV}) together with the boundary
conditions (\ref{FBC}) and (\ref{FCAC}). We have calculated the
frequencies of the $f$-modes and the higher fluid modes $p_1$ and
$p_2$. All of the presented dependences  are shown up to the
maximum mass for the corresponding parameters where the solutions
become unstable \cite{Horvat}.

\begin{figure}
\begin{center}
\includegraphics[width=0.6\textwidth]{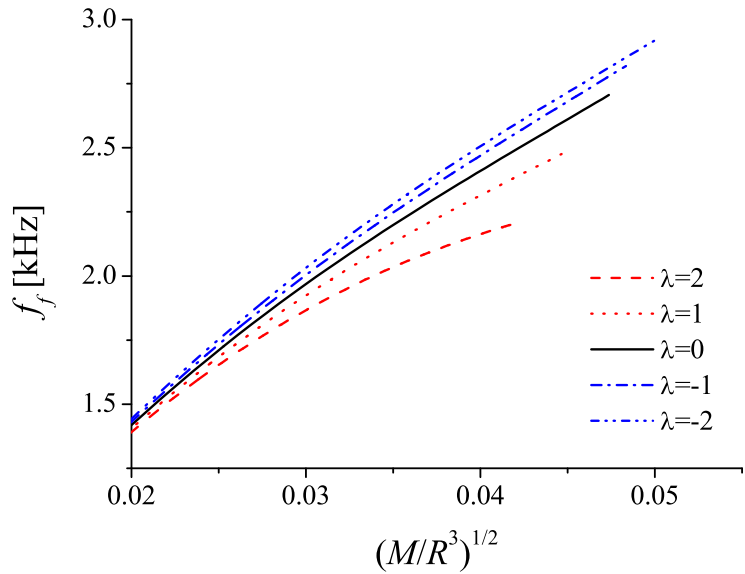}
\end{center}
\caption{The $f$-mode frequency as a function of the square root of the average density $\sqrt{M/R^3}$. The results
for several values of the parameter $\lambda$ are shown.} \label{fig:f(sqrtMR3)_fmode}%
\end{figure}

An empirical dependence between the $f$-mode frequencies and the
average density was found in \cite{Kokkotas01,Andersson98} for the
case of isotropic neutron stars and it is interesting to see if it
changes in our case. The $f$-mode oscillation frequencies as a 
function of the square root of the average density are presented in Fig.
\ref{fig:f(sqrtMR3)_fmode} for several values of the parameter
$\lambda$. The graph shows that the dependence does not change
significantly for small values of the average density. Only for
large values of the average density and for large absolute values of
$\lambda$, the deviation from the isotropic neutron stars  (i.e.
when $\lambda=0$) is more significant. But still the uncertainties
in obtaining the coefficients in the empirical dependence in
\cite{Andersson98}, which come from varying the equation of state,
are comparable with the deviation due to the anisotropic pressure.

\begin{figure}
\includegraphics[width=0.48\textwidth]{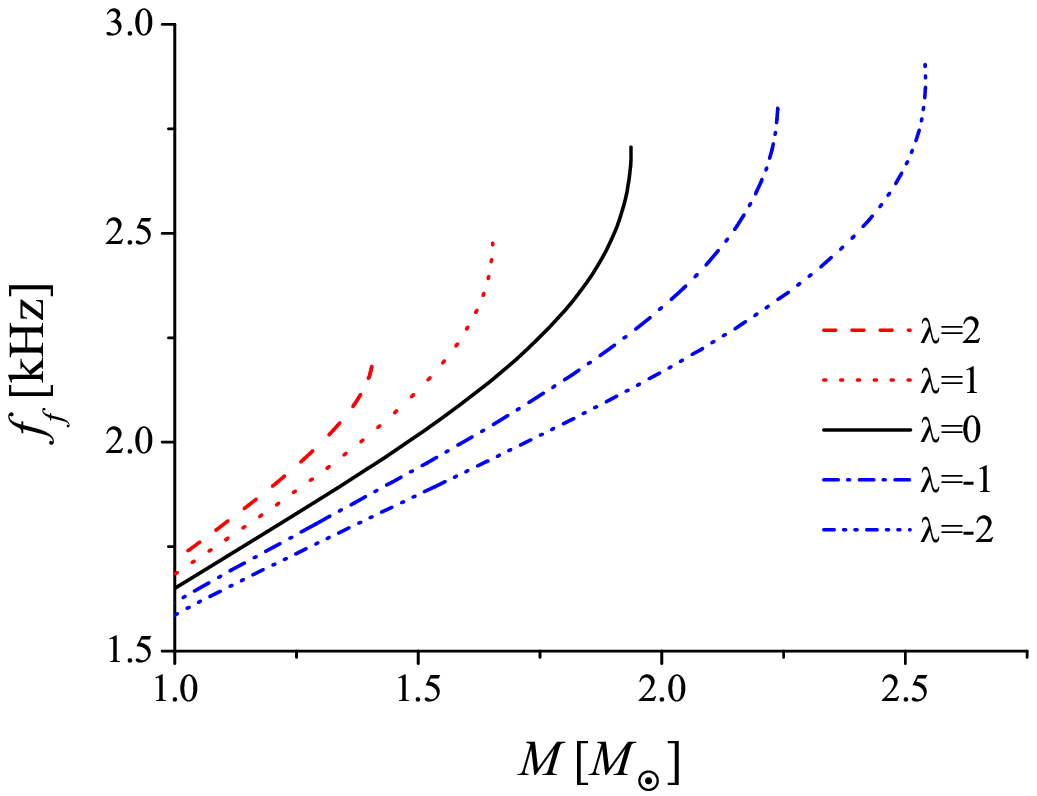}
\includegraphics[width=0.48\textwidth]{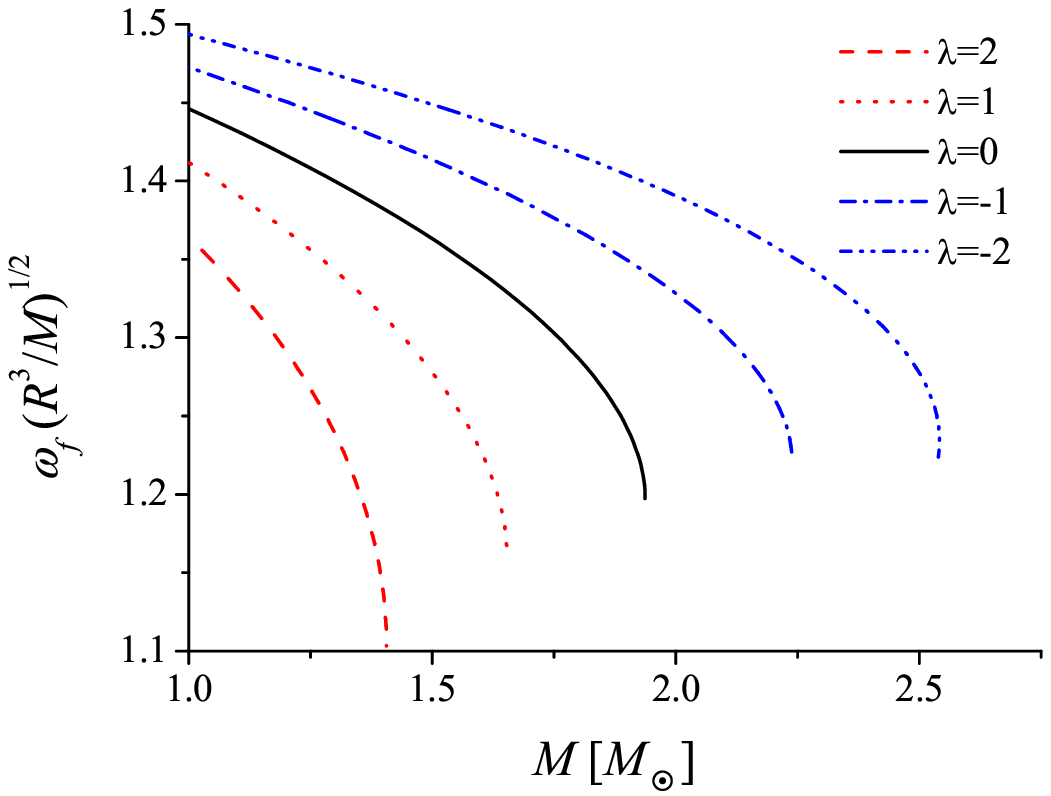}
\caption{The frequency $f$ as a function of the mass $M$ (left panel) and the normalized frequency $\omega$ as a
function of $M$ (right panel) for the $f$ mode.  The results
for several values of the parameter $\lambda$ are shown.} \label{fig:Omega(M)_fmode}%
\end{figure}%

\begin{figure}
\includegraphics[width=0.48\textwidth]{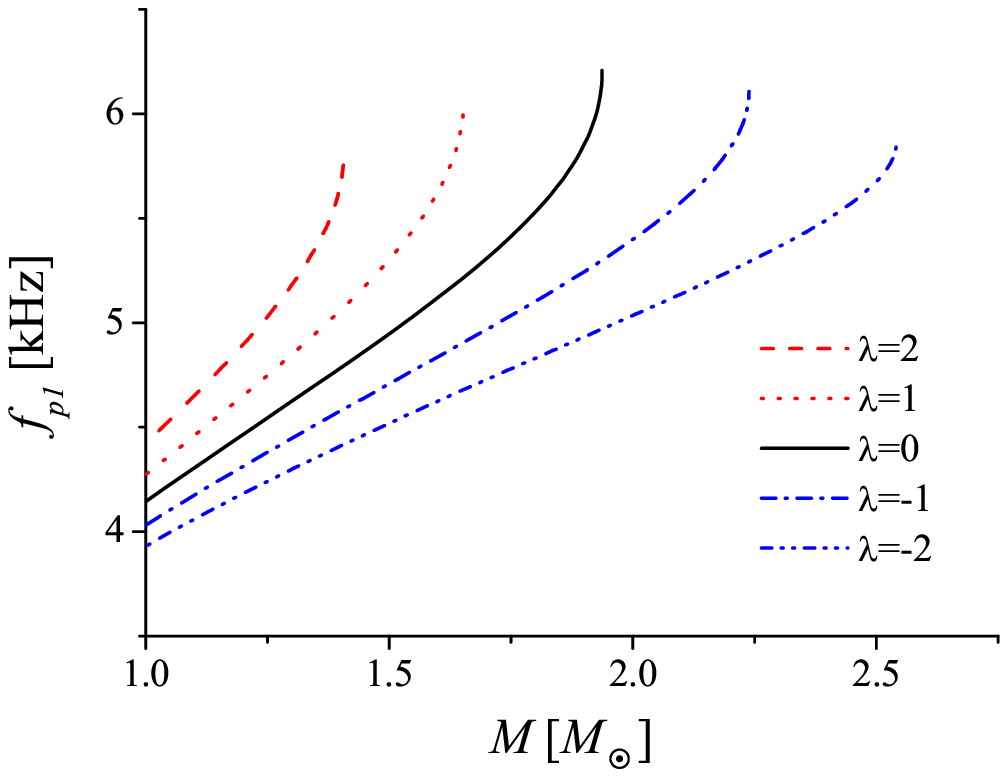}
\includegraphics[width=0.48\textwidth]{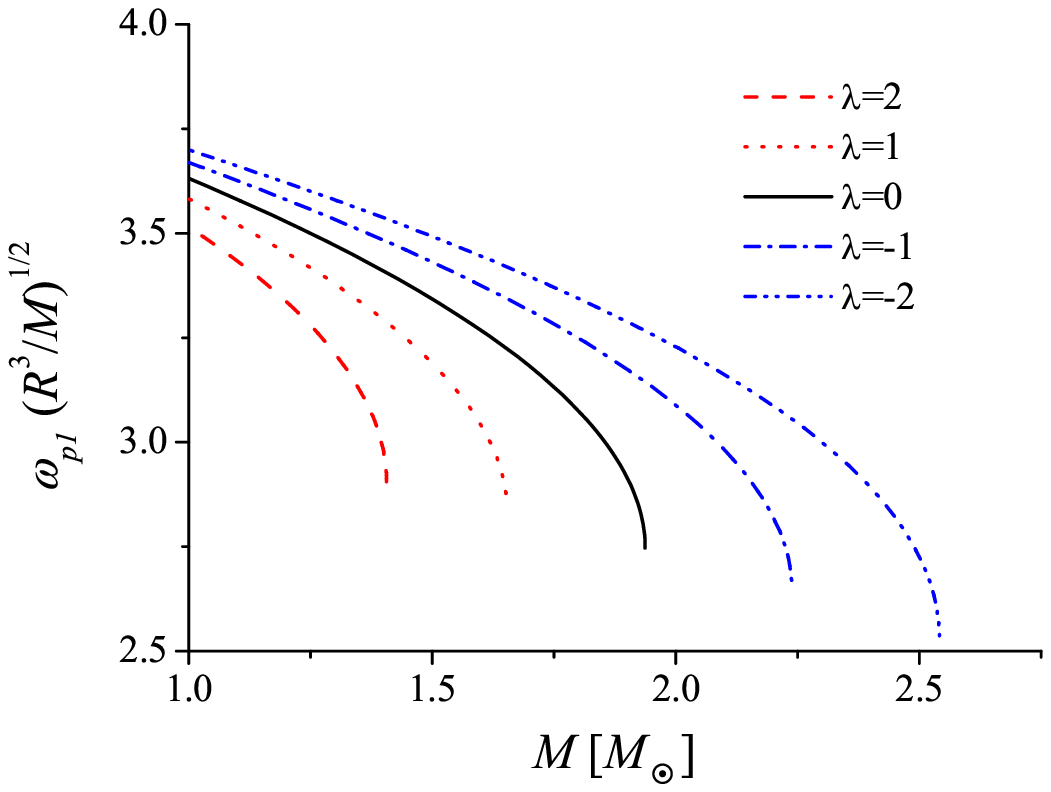}
\caption{The results for the $p_1$-mode of the same solutions as shown on figure \ref{fig:Omega(M)_fmode}.} \label{fig:Omega(M)_p1mode}%
\end{figure}%

The frequency $f$ and the normalized frequency $\omega \sqrt{R^3/M}$
as a function of the mass are shown in Figs.
\ref{fig:Omega(M)_fmode} and \ref{fig:Omega(M)_p1mode} for the $f$
and the $p_1$ modes. Depending on the sign of $\lambda$ the
frequencies can be larger or smaller than in the case of isotropic
neutron stars. As we can see the frequencies can change considerably
when we increase the absolute value of $\lambda$. Also for fixed
value of $\lambda$ the differences with the isotropic neutron stars
are bigger for larger masses because in this case the compactness
$\mu$, which enters  the EOS for the anisotropic pressure
(\ref{eq:EOS_sigma}), is larger. Therefore for large absolute values
of $\lambda$ and for large masses the oscillation frequencies can
differ significantly from the isotropic neutron star case.

\begin{figure}
\begin{center}
\includegraphics[width=0.6\textwidth]{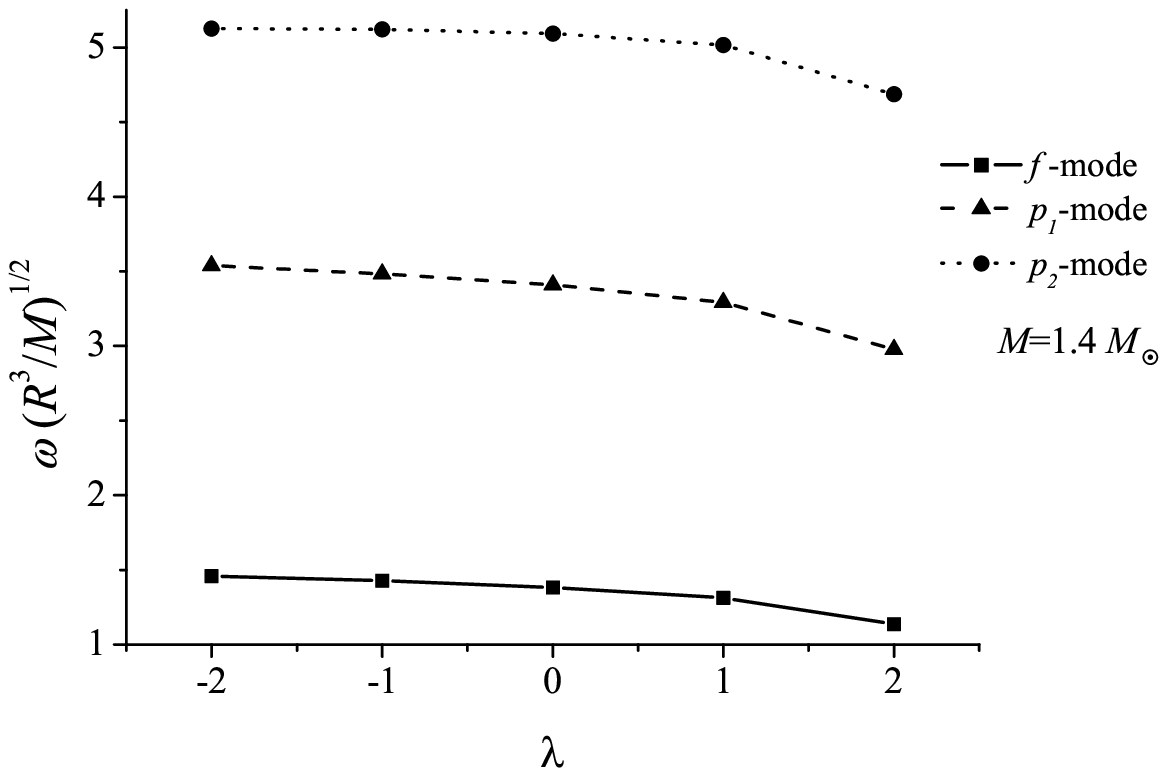}
\end{center}
\caption{The normalized frequency $\omega$ as a function of $\lambda$ for fixed value of the mass $M=1.4 M_{\bigodot}$.
The
results for the $f$, $p_1$ and $p_2$ modes are shown.} \label{fig:Omega(kAnis)_M1.4}%
\end{figure}%

The normalized frequencies of the $f$, $p_1$ and $p_2$ modes as a
function of $\lambda$ are shown in Fig. \ref{fig:Omega(kAnis)_M1.4}
where the mass $M=1.4 M_{\bigodot}$ is the same for all  the
solutions. As we can see the changes in the frequencies as we vary
$\lambda$, are similar for all the modes. This behavior is
qualitatively different from some of the alternative models of
neutron stars where the oscillations frequencies vary more
significantly as a function of the corresponding parameter, for the
higher fluid modes \cite{Sotani04}--\cite{Yazadjiev11}.

\begin{figure}
\begin{center}
\includegraphics[width=0.6\textwidth]{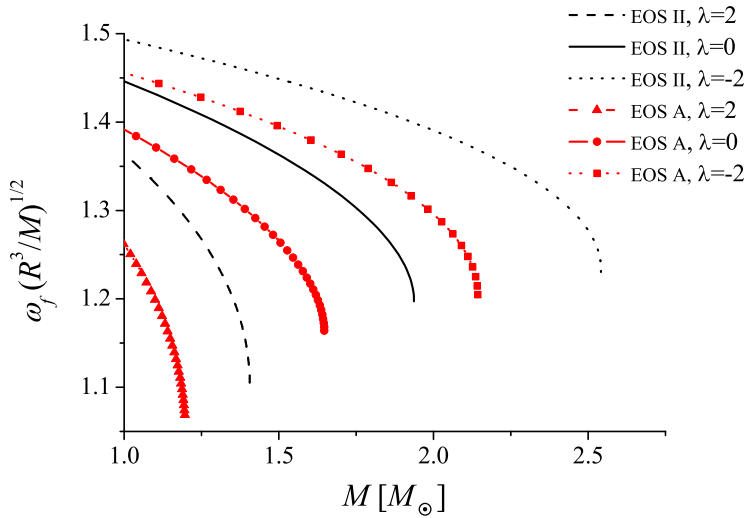}
\end{center}
\caption{The normalized frequency $\omega$ as a function of $M$ for the $f$ mode.  The results for different values of
$\lambda$ and for two polytropic equations of state of the radial pressure are shown -- a soft equation of state EOS A
and a stiff equation of state EOS II.} \label{fig:Omega(M)_fmode_EOS}%
\end{figure}

It is interesting to compare the effects on the oscillation spectrum
caused by varying the anisotropic pressure and by changing the
equation of state of the radial pressure. As we said before the
dependence between the $f$-mode frequencies and the average density
does not change much when we vary the equation of state. The same is
true also when we vary the anisotropic pressure. But the normalized
frequency $\omega$ as a function of the mass changes significantly
when we vary $\lambda$ and the equations of state. This can be seen
on Fig. \ref{fig:Omega(M)_fmode_EOS} where the results for the
$f$-mode oscillation frequencies are presented for two equations of
state of the radial pressure and for several values of $\lambda$.
The EOS II is the standard polytropic equation of state which we
used up to now with $\Gamma=2.34$ and $K=0.0195$. The EOS A is again
a polytropic equation of state where the coefficients $\Gamma=2.46$
and $K=0.00936$ are obtained when fitting the tabulated data for EOS
A \cite{Arnett77} and EOS II is stiffer than EOS A. As we can see
the presence of an anisotropic pressure changes the frequencies in a
similar way as changing the EOS, and more precisely positive values
of $\lambda$ lead to frequencies similar to a softer EOS, and
negative values of $\lambda$ lead to frequencies similar to a
stiffer EOS. Thus the oscillation spectrum of neutron stars with
anisotropic pressure can mimic to a certain extent the oscillation
spectrum of neutron stars with softer/stiffer equation of state. But
still if we consider strong anisotropic pressure the frequencies can
change a lot which is hard to be achieved by the standard nuclear
equations of state. Thus observing more than one fluid mode of a
neutron star can help us to prove or at least set limits on the
possible existence of anisotropic pressure in the neutron stars.

\section{Conclusions}
In the present paper we study how  the possible existence of an
anisotropic pressure inside a neutron star can change the
oscillation spectrum. As a first step we examine the oscillations in the
Cowling approximation where the metric is kept fixed and within this
approximation the perturbation equations for the anisotropic neutron
stars are derived.

The background solution are obtained numerically by solving the reduced field equations where the equation of state for
the radial pressure is polytropic and we use a quasilocal equation of state for the anisotropic pressure
\cite{Horvat}. The properties of the obtained solutions can differ significantly from the isotropic neutron stars.

The oscillation spectrum of the anisotropic neutron stars is
obtained when solving the perturbation equations with the
appropriate boundary condition and the results for the $f$-mode and
the higher fluid modes are obtained. It turns out that the
dependence between the $f$-mode frequencies and the average density
which was obtained in \cite{Andersson98} does not change much in the
presence of an anisotropic pressure. The effect of the anisotropy is
more evident on other dependences, for example, the normalized
frequency as a function of the mass changes considerably for
anisotropic stars. Thus the observation of more than one fluid mode
can serve as a test for the existence of an anisotropic pressure in
the neutron stars.

We have also compared the effect on the oscillation spectrum caused
by the anisotropic pressure  and by changing the equation of state
of the radial pressure. It turns out that for negative values of the
anisotropic pressure, i.e. for $\lambda<0$, the changes in the
frequencies are similar to what we will obtain in the case of
isotopic neutron star with a stiffer equation of state, and for
$\lambda>0$ the results are similar to the case of an isotopic
neutron star with a softer equation of state. A more detailed
analysis, for example if we drop the Cowling approximation or if we
consider rotating solutions, may show more differences between the
oscillation spectrum of  isotropic and  anisotropic neutron stars
and we plan to make such a study in the future.

It would be also interesting to check how a change in the equation
of state of the anisotropic pressure influences the results because
the choice of the quasilocal EOS (\ref{eq:EOS_sigma}) is by no
means the only possible one. Up to now, however,  little is know
about the EOS of the anisotropic pressure and that is why further
studies on the possible quasilocal equations of state and their
effect on the stellar structure and oscillations are needed.

\vspace{1.5ex}
\begin{flushleft}
\large\bf Acknowledgments
\end{flushleft}

The authors would like to thank K. Kokkotas for reading the manuscript and for the valuable suggestions.  S.Y. would
like to thank the Alexander von Humboldt Foundation for the support, and the Abteilung Theoretische Astrophysik
T\"ubingen for its kind hospitality. D.D. would like to thank the DAAD for the support and the Abteilung Theoretische
Astrophysik T\"ubingen for its kind hospitality. D.D. is also supported by the German Science Council (DFG) via
SFB/TR7. This work was also supported in part by  the Bulgarian National Science Fund under Grants DO 02-257 and
DMU-03/6.

\end{document}